\documentclass{article}

\usepackage{PRIMEarxiv}

\usepackage[utf8]{inputenc} 
\usepackage[T1]{fontenc}   
\usepackage{url}            
\usepackage{amsfonts}       
\usepackage{nicefrac}       
\usepackage{microtype}     
\usepackage{fancyhdr}       

\usepackage{setspace}
\usepackage{xcolor}
\definecolor{tikz_color}{rgb}{0.55, 0.57, 0.67} 
\definecolor{tikz_white}{rgb}{0.90, 0.89, 0.89} 

\usepackage[inline]{enumitem}

\usepackage{longtable}
\usepackage{caption}
\usepackage{bbm, dsfont}
\usepackage{setspace}
\usepackage{booktabs}  
\usepackage{multirow}
\usepackage{makecell}
\usepackage{adjustbox}
\usepackage[flushleft]{threeparttable}
\usepackage{float}

\usepackage{tabularx}
\usepackage{amsmath}
\usepackage{amssymb}
\usepackage{amsbsy}
\usepackage{amsthm}
\usepackage{bm}
\usepackage{mathtools}

\usepackage{tikz}
\tikzset{>=latex}
\tikzset{edge/.style={thick}}
\usepackage{circuitikz}

\newcommand{\myfrac}[3][3pt]{\genfrac{}{}{}{}{\raisebox{#1}{$#2$}}{\raisebox{-#1}{$#3$}}}
\newcommand\numberthis{\addtocounter{equation}{1}\tag{\theequation}}

\newcolumntype{C}{>{\centering\arraybackslash}X}

\usepackage[authoryear]{natbib}
\usepackage[hidelinks]{hyperref}

\title{Relational Event Modeling}

\author{
  Federica Bianchi, Edoardo Filippi-Mazzola, Alessandro Lomi, and Ernst C. Wit \\
  Universit{\`a} della Svizzera italiana\\
  Lugano, Switzerland\\
  \texttt{\url{federica.bianchi@usi.ch}} \\
}

\begin{document}
\maketitle

\begin{abstract}
Advances in information technology have increased the availability of time-stamped relational data such as those produced by email exchanges or interaction through social media. Whereas the associated information flows could be aggregated into cross-sectional panels, the temporal ordering of the events frequently contains information that requires new models for the analysis of continuous-time interactions, subject to both endogenous and exogenous influences. The introduction of the \emph{Relational Event Model} (REM) has been a major development that has led to further methodological improvements stimulated by new questions that REMs made possible. In this review, we track the intellectual history of the REM, define its core properties, and discuss why and how it has been considered useful in empirical research. We describe how the demands of novel applications have stimulated methodological, computational, and inferential advancements.
\end{abstract}


\section*{Introduction}
\label{sec:introduction}

Statistical models for social and other networks are receiving increased attention not only in specialized field journals such as \emph{Network Science} or \emph{Social Networks}, but also in prominent interdisciplinary science journals such as \emph{Science} \citep{borgatti2009network, butts2009revisiting}, \emph{PNAS} \citep{stadtfeld2019integration}, and \emph{Science Advances} \citep{elmer2020depressive}.

Attention to statistical models for networks is on the rise also in well-established generalist statistics journals such as, for example, the \emph{Journal of the American Statistical Association} \citep{hunter2008goodness}, the \emph{Journal of Applied Statistics} \citep{snijders2010maximum}, \emph{Statistical Science} \citep{schweinberger2020exponential}, and the \emph{Journals of the Royal Statistical Society (series A, B, and C)} \citep{fienberg2012brief, krivitsky2014separable, gile2017analysis, vinciotti2017preface, koskinen2023multilevel}. The \emph{Annual Review of Statistics and Its Applications} itself has recently demonstrated considerable interest in models for social networks by publishing comprehensive and up-to-date reviews on two popular classes of statistical models: \emph{Exponential Random Graph Models} (ERGMs) \citep{amati2018social} and \emph{Stochastic Actor-Oriented Models} (SAOMs) \citep{snijders2017stochastic}.

Since the publication of these reviews, the increasing availability of time-stamped data resulting from innovation in data production, collection, storage and retrieval technologies has shown that network data samples collected at fixed time intervals are likely to miss fundamental differences in the time scales over which relational processes unfold \citep{golder2007rhythms}. Computer-mediated communication \citep{lerner2023relational}, sociometric badges \citep{wu2008mining, stehle2011high, elmer2020depressive}, electronic trading platforms \citep{zappa2021markets}, on-line interaction logs \citep{tonellato2023microstructural}, and video recordings \citep{pallotti2022lost}, are just some of the new data-generating technologies capable of producing large quantities of relational event data connecting sender and receiver units. 

During the same period, studies based on event-oriented designs have become also increasingly common. While the empirical opportunities offered by relational event data have long been acknowledged by students of social networks \citep{freeman1987cognitive, marsden1990network, borgatti2009network}, statistical models affording a degree of temporal resolution consistent with the frequency of observed social interaction \citep{butts2009revisiting} have become available only during the last fifteen years \citep{butts2008relational}. 

Time scales vary considerably based on interaction settings. High-frequency transactions in financial markets \citep{bianchi2022from} occur within seconds or fractions of seconds, while communication in emergency situations \citep{butts2008relational, renshaw2022modeling} may take several minutes. Email exchange \citep{perry2013point} can extend over hours, whereas interaction generated by more complex forms of social coordination among corporate actors may become observable only over days or even weeks \citep{amati2019some}. In these various cases, aggregating time-stamped relational event data into network ties defined over conventional, or convenient, time periods, is unlikely to afford high fidelity representations of the underlying interaction processes \citep{tuma1984social}.

Perhaps the main motivation that inspired the development of the \emph{Relational Event Model} (REM) proposed by \citet{butts2008relational}, was to provide a general analytical framework where \emph{sequential ordering} and \emph{timing} replace \emph{concurrency} and \emph{temporal aggregation} of network edges as the ``dominant concepts of phenomenal concern'' \citep[p.192]{butts2008relational} in the analysis of social interaction. This involves deriving, specifying and estimating statistical models capable of assimilating and analyzing complex relational data without altering --- through time aggregation --- the natural time structure, and sequential ordering of observed social interaction. Conversation \citep{gibson2005taking}, communication \citep{pilny2017adapted}, market exchange \citep{lomi2021time}, and other, more complex, forms of social coordination \citep{lerner2020free} can be understood only with reference to the timing and sequential order of the relational events of interest \citep{abbott1992causes}, which contain important information that is typically lost when time-stamped events are aggregated into binary network ``ties'' \citep{pallotti2022lost}.

Since its introduction, the REM has been significantly refined and adapted to an ever-increasing diversity and sophistication of emerging empirical problems \citep{butts_lomi_snijders_stadtfeld_2023}. This review piece provides an opportunity to position relational event modeling in the broader context of statistical models for network science, and assess the current state of field, incorporating a broad review of contemporary methodological, computational, and inferential developments in this class of statistical models for directed social interaction.

The paper is organized into eight main sections. Section~\ref{sec:historical-context-REMs} traces the intellectual origins and historical context of REMs. Section~\ref{sec:relational-data_study-design} identifies the observation plans and empirical research design elements typically associated with event-oriented studies of network dynamics. Section~\ref{sec:types_REMs} surveys the available classes of REMs developed, at least in part, in response to specific problems with no satisfactory modeling solutions. Section~\ref{sec:estimation-computation} is dedicated to issues of empirical model specification and estimation. Section~\ref{sec:emprical-applications} examines the broad area of applications where REMs seem to have found fertile ground for development. Section~\ref{sec:open-issues-challenges} reviews the main challenges and open issues that orient current research. Section~\ref{sec:summary_conclusions} concludes with a summary and a discussion of the main promises of relational event modeling.
\section{Historical Context of Relational Event Models}

\label{sec:historical-context-REMs}

Contemporary statistical models for social and other kind of networks rely heavily on the formalism of graph theory \citep{butts2009revisiting} --- an inheritance left, in part, by earlier network models developed in sociometry \citep{moreno1934shall, jennings1948sociometry}, and within the structural tradition of social and cultural anthropology \citep{white1963anatomy,levi1971elementary,hage1979graph,barnes1983graph, hage1984structural}. Concepts such as those of ``degree,'' ``path distance,'' ``reciprocity,'' and ``transitive closure'' that are central in contemporary statistical models for networks \citep{snijders2001statistical, amati2018social} are firmly rooted in the mathematical representation of networks as graphs \citep{barabasi2016network}. 

The graph-theoretic formalism that inspired early Markov random graph models \citep{holland1977dynamic, holland1981exponential, frank1986markov} may be considered to be the intellectual antecedent of contemporary statistical models for social networks \citep{wasserman1996logit, snijders2001statistical, snijders2006new}. It is only in relatively recent times that the limitations in representing networks of social relations as graphs have started to become apparent. REMs entail an alternative understanding of social relations as emergent from sequences of relational events connecting a sender behavioral unit to one or more receivers \citep{butts2008relational, butts2009revisiting, perry2013point, lerner2023relational}.

REMs provide a framework for analyzing and making inferences about the relational processes and dynamics in complex social systems. They are designed to capture the patterns of dependence in the occurrence and timing of relational events, such as communications, transactions, or social interactions, within a network. 

Formally, REMs are rooted in event history models \citep[see, e.g.,][for a review]{keiding2014event}, expressing the hazards of an event to occur, as a function of the history of previous events, as well as potentially additional nodal and relational attributes \citep{butts2008relational}. REMs allow to study how events unfold over time, how they are influenced by various attributes and how they, in turn, affect the structure and evolution of the network.

The exact hazards of a REM may follow different functional forms and specifications. In many instances, the hazard is assumed to be piecewise constant, resulting in waiting times that are conditionally exponentially distributed. However, the precise definition of hazards will depend on the research questions and the empirical context, and these choices have implications for computational complexity, model fit, and the interpretation of the model \citep{butts2017relational, schaefer2017modeling, stadtfeld2017dynamic}.

Two other modeling frameworks are appropriate for the analysis of longitudinal networks, the temporal ERGM and the SOAM. The formulation of the ERGM \citep{robins2007introduction} may be understood in terms of the conditional probability of a network edge, given the relationships among all other network edges \citep{wasserman1996logit, snijders2006new}. In the ERGM the edges connecting pairs of senders and receivers are assumed to be interdependent, and such interdependencies can be captured by local configurations of network ties such as triangles or star-shaped structures \citep{robins2007introduction}. More specifically, the existence of a tie in a social network is conditionally independent of ties that are far distant in space \citep{frank1986markov}. Temporal extensions of the ERGM \citep{hanneke2010discrete, krivitsky2014separable} have been introduced to model network dynamics over time.

The SAOM \citep{snijders1996stochastic, snijders2001statistical, snijders2005models,snijders2017stochastic}, is a probability model evaluating change in network ties (relational states) observed in adjacent time periods. While the SAOM takes into account relational states observed at multiple time points, it operates under the assumption that ties change continuously over time through a series of micro-steps occurring between observations. Each micro-step involves at most one alteration \citep{holland1977dynamic} in the network with social actors choosing what ties to alter based on a multinomial choice probability model \citep{mcfadden1973conditional} accounting for the local configurations in which potential ties are embedded.
\section{Relational Data and Study Design}

\label{sec:relational-data_study-design}

The relational event modeling framework has been introduced to analyze the dynamics underlying social networks, with a focus on understanding the complex interactions and dependencies between behavioral units over time. REMs are flexible because they integrate temporal, structural, and attributional aspects of the data within a single framework. This flexibility sustains specification of very general, yet detailed models of network evolution \citep{brandes2009networks}.

\subsection{Network Structure}
\label{subsec:network-structure}

The building block of the REM is the relational event, defined as an interaction initiated by a sender unit and directed toward one or more targets \citep{butts2008relational}. Examples include hospitals sharing patients \citep{vu2017relational, amati2019some}, credit institutions exchanging financial assets \citep{lomi2021time, zappa2021markets, bianchi2022from}, students telephoning each other \citep{stadtfeld2017interactions}, people interacting face-to-face \citep{elmer2020depressive} and in social groups \citep{hoffman2020model}, or employees exchanging emails \citep{quintane2013short, perry2013point}. 

REMs can also model relations between units belonging to different classes in the context of more general bipartite processes. Examples include non native species invading new spatial niches \citep{juozaitiene2023analysing}, editors modifying Wikipedia pages \citep{lerner2017third}, computer developers fixing software problems \citep{tonellato2023microstructural}, and political actors supporting or rejecting proposed legislation \citep{brandenberger2018trading, haunss2022multimodal}. 

Although relational events are typically defined between a single sender and a single receiver, generalizations are possible where a sender may reach multiple receivers simultaneously. This situation is common in technology-mediated communication \citep{lerner2023relational}, citation networks \citep{lerner2023micro, filippi2023dream}, and social gatherings attended by many participants \citep{lerner2021dynamic}.

\subsection{Sampling and Recording}
\label{subsec:sampling_recording}

Event network studies typically look at social units interacting within a bounded environment. Examples of geographical areas consist of student houses \citep{stadtfeld2017interactions} or hospitals within geographical regions \citep{vu2017relational, zachrison2022influence}, whereas examples of temporal boundaries involve interaction between emergency teams during different phases of a crisis \citep{butts2008relational}, or hospitals transferring patients in specific days of the week \citep{amati2019some}. Relational events taking place outside the main observation domain are usually not recorded, so the social networks reconstructed from streams of relational events are often incomplete. Unlike studies of statistically independent observations, network boundary specification might affect the validity, robustness, and replicability of the results \citep{laumann1989boundary}.
\section{Specifications of Relational Event Models}
\label{sec:types_REMs}

The units of analysis in the REM are the edges connecting individual pairs of senders and receivers. Those edges are typically stored in tuples $(t, s, r)$, where $s$ is the sender, $r$ is the receiver, and $t$ is the time of the relational event connecting $s$ to $r$. At its core, the REM is defined as a point process for directed pairwise interactions that, in turn, are modeled through their rate function $\lambda$. The model assumes that $\lambda$ may depend upon sender, receiver, past event history, and/or exogenous covariates.

\begin{table}[htbp]
\caption*{Notation}
\label{tab:notation}
\centering
\begin{tabularx}{\textwidth}{cXl}
\toprule
Notation & Meaning \\ 
\midrule
$(t,s,r)$ & A relational event in which sender $s$ interacts with receiver $r$ at time $t$  \\
$\lambda_{sr}(t)$ & Rate/hazard at which sender $s$ contacts receiver $r$ at time $t$ \\
$\mathcal{H}_t$ & Filtration of the process, containing all information about relational events until time $t$ \\
$L,L_P$ & Likelihood and partial likelihood \\
$\mathcal{R}(t)$ & Risk set at time $t$, consisting of all relational events that can happen at that time  \\
$\widetilde{\mathcal{R}}(t)$ & Sampled risk set at time $t$, consisting of the event $(s, r)$ at time $t$ and a set of sampled non-events \\
$x_{sr}(t)$ & Dyadic covariate(s) with corresponding effect(s) $\beta$ \\
$z_{sr}(t)$ & Dyadic covariate(s) with corresponding random effect(s) $\gamma$  \\
$a$ & Alter, i.e., an individual different from sender or receiver \\ 
\bottomrule
\end{tabularx}
\end{table}

\subsection{Types of Relational Event Models}
\label{subsec:REMs-types}

We consider a fixed time interval $[0,T]$, with $0<T<\infty$, in which \emph{events} occur. Events are defined as time-stamped interactions between senders and receivers. Both the set of senders $\mathbf{S}$ and receivers $\mathbf{R}$ are assumed to be finite. For \emph{one-mode networks} the set of senders and receivers overlap, $\mathbf{S} = \mathbf{R}$, whereas for \emph{two-mode} or \emph{bipartite networks} they are distinct. The relational event process is a marked point process \citep{daley2003introduction, cressie2015statistics} based on more general approaches \citep{borgan1995methods, aalen2008survival} for event history sequences $\left\{(t_i, s_i, r_i): ~i \geq 1, s_i \subset \mathbf{S}, r_i \subset \mathbf{R} \right\}$, and defined on a probability space $(\Omega,\mathcal{F}, P)$ adapted to the filtration $\mathcal{H}_t$, consisting of the history of process \citep{andersen1993statistical}. In principle, the marks $(s, r)$ can be individuals or sets of senders and receivers. 

Associated with this marked point process, we define a multivariate counting process $N$, whose components $N_{sr}$ record the number of directed interactions between $s$ and $r$,
\[ N_{sr}(t) = \sum_{i\geq 1} \mathbbm{1}_{\left\{t_i\leq t;~ s_i=s;~ r_i=r\right\}}. \] 
According to the Doob--Meyer decomposition theorem \citep{meyer1962decomposition}, there exists a predictable process $\Lambda_{sr}$ and a residual martingale process $M_{sr}$, such that the counting process can be written as
\[ N_{sr}(t) = \Lambda_{sr}(t) + M_{sr}(t).\]

The aim of the REM is to describe the structure of the predictable cumulative hazard process $\Lambda_{sr}$. By assuming that the counting process is an inhomogeneous Poisson process, we can write the cumulative hazard as
\[ \Lambda_{sr}(t)= \int_0^t \lambda_{sr}(\tau)~d\tau,\]
where $\lambda_{sr}$ is the hazard function of the relational event $(s, r)$. The general REM is defined as
\begin{equation}
\lambda_{sr}(t) = \mathbbm{1}_{\{(s, r)\in \mathcal{R}(t) \}} \: \lambda_0(t) \: \psi \Bigl\{ \beta(t), x_{sr}(t)\Bigr\},
\label{eq:general-rem}	
\end{equation}
where $\lambda_{sr}(t)$ is only non-zero if the event $(s, r)$ is contained in the \emph{risk set} $\mathcal{R}(t)$ of possible events at time $t$, $\lambda_0(t)$ is the baseline hazard function unrelated to $(s, r)$, $\psi$ is a \emph{relative risk function} as defined in \citet{thomas1981general}, $x_{sr}(t)$ is the $\mathcal{H}_t$ measurable set of endogenous and exogenous (possibly) time-varying variables, and $\beta(t)$ are the effect sizes. The function $\psi$ is positive and, for identifiability, normalized by setting $\psi(\beta, 0)=1$. 

The relational event literature has focused on the exponential risk function $\psi(\beta, x) = \exp(\beta^{\top} x)$. Under the conditional independence framework \citep{besag1974spatial, besag1975statistical}, network statistics are, by construction, dependent through space and time. Accordingly, the propensity of pairs of senders and receivers to mutually connect depends on which relational events they have sent or received in the past, their order in time, if not their exact timing, and on exogenous factors like nodal or network attributes. 

\subsubsection{The Original Relational Event Model}
\label{subsubsec:original_REM}

\cite{butts2008relational} introduced the REM as a piece-wise homogeneous Poisson process, whereby the rate function was specified as
\begin{equation}
\label{eq:model_REM}
\lambda_{sr} (t) = \lambda_{0} \: \exp \Bigl\{ \beta^{\top} x_{sr}(t) \Bigr\} \mathbbm{1}_{\left\{(s, r)\in \mathcal{R}(t) \right\}},
\end{equation}
where $x_{sr}(t)$ is the collection of sufficient network statistics associated with the parameter vector $\beta \in \mathbb{R}^{p}$. The covariates in $x_{sr}(t)$ can depend on nodal characteristics, such as the age of $r$ or the age difference between the pair $(s, r)$, and also on the history of the process. For instance, the covariates $x_{sr}(t)$ can include a count of the number of past events between $(s, r)$ before time $t$. 

In the original formulation, the baseline rate of the interaction process $\lambda_{0}$ is assumed constant. This means that the waiting times between events is exponentially distributed, and once an event takes place, the context of the action is changed, and all the possible relational event waiting times are restarted.

\subsubsection{Weighted and Signed Networks}
\label{subsubsec:weighted-signed-networks}

One of the first extensions of the original REM involved geopolitical conflicts and cooperation \citep{brandes2009networks}. The authors highlighted that each interaction event between countries, international organizations, or ethnic groups could be given either a positive or negative weight according to the degree of cooperation or hostility between countries. 

In this extended relational event definition $(t,s,r,w)$, the interaction weight $w$ is modeled alongside the event $(t, s, r)$ as $p\left\{(t, s, r, w)~|~\mathcal{H}_{t-}\right\} = p\left\{(t, s, r)~|~\mathcal{H}_{t-}\right\} \times p\left\{w~|~(t, s, r),\mathcal{H}_{t-}\right\}$. The second term models the interaction weights depending on which countries are interacting and how they interacted in the past, whereas the first term is the usual REM likelihood, which is also allowed to depend on past interaction weights included in $\mathcal{H}_{t-}$. 

Subsequent versions of the REM have accommodated weighted network statistics \citep{amati2019some, bianchi2022from} associating a decay function $f\left(t, t_{i}, \alpha \right)$ to each past event $(s_i,r_i)$ where $t$ is the current time, and $\alpha$ a decay parameter. In its simplest formulation, the temporal relevance of an event is assumed to decrease according to a power law, $f \left( t, t_{i}, \alpha \right) = \left( t - t_{i} \right)^{-\alpha}$, though other specifications may be used instead \citep{lerner2020reliability}.

\subsubsection{Multicast or Polyadic Interaction}
\label{subsubsec:multicast-polyadic-networks}

Multicast (or ``polyadic'') interaction occurs when a single event links an individual sender to multiple receivers simultaneously. The possibility of one-to-many interaction was explicitly recognized by \citet{dubois2013hierarchical} in their model of teacher-students interaction in a classroom context. In the same year, \citet{perry2013point} provided a more general formulation specification of the intensity function for a sender $s\in\mathbf{S}$ addressing a set of receivers $R = \{ r_{1}, \dots, r_{m} \} \subset \mathbf{R}$, taking
\begin{equation}
\label{eq:model_perry2013point}
\lambda_{sR}(t) = \lambda_{0s}(t;\lvert R \rvert) \: \exp \left\{ \beta_{0}^{\top} \sum_{r \in \mathcal{R}_{s}(t)} x_{sr} \left(t \right) \right\} \prod_{r \in R} \mathbbm{1}_{\left\{ r \in R_{s}(t) \right\}}. 
\end{equation}
\noindent
The event rate function involves two types of stratification, discussed in greater detail in Sections~\ref{subsubsec:separable-intensity-functions} and \ref{subsubsec:stratification}, which explains the sender-specific definition of the risk set $\mathcal{R}_s(t)$ and baseline hazard, which also depend on the size of the receiver set. The network statistics are defined as the sums of the individual dyadic statistics $x_{sR}(t) = \sum_{r \in R} x_{sr}(t)$. 

\citet{lerner2023relational} introduced the \emph{Relational Hyperevent Model} (RHEM) by generalizing the definition of multicast network statistic. A hyperedge covariate $x_{sR}$ is a function of the sender and the \emph{entire set} of receivers that cannot, necessarily, be decomposed into the sum of dyadic covariates. An example of hyperedge covariate is \emph{inertia}, the tendency toward repeating past relational events, defined as
\[ 
x_{sR}^{\mbox{\scriptsize inertia}}(t) = \sum_{t_i<t} f(t ,t_i,\alpha) \mathbbm{1}_{\left\{s_i=s,R_i=R\right\}}. 
\]
In email exchange, for instance, the existence of a mailing list may result in a moderator communicating with exactly the same group of individuals. Instead, \emph{unordered repetition} captures interaction within a stable set of actors with turn-taking among the senders \citep{gibson2005taking}. Replacing dyadic covariates with their hyperedge counterparts not only provides a richer collection of network effects but also improves model fit \citep{lerner2023relational}. 

Alternative approaches for modeling polyadic interactions have been proposed. \citet{kim2018hyperedge} introduced the \emph{Hyperedge Event Model}, assuming that dyadic intensities stochastically determine the sender of the next event and its receiver set. \citet{mulder2021latent} introduced a latent variable model whereby all potential receivers are assigned to the receiver set on the basis of a suitability score depending on the sender and receiver-specific characteristics.

\subsubsection{Separable Intensity Functions}
\label{subsubsec:separable-intensity-functions}

\citet{stadtfeld2017dynamic} developed the \emph{Dynamic Network Actor Model} (DyNAM), an actor-oriented model for relational event data built upon the same paradigm introduced by \citet{snijders1996stochastic, snijders2017stochastic}. The distinctive feature of the DyNAM is that it explains social interaction in terms of ``individuals' preferences, available interaction opportunities, and individuals' perception of the social network they are embedded in'' \citep[p.318]{stadtfeld2017interactions}. Accordingly, the inferential framework of the DyNAM consists of modeling a composite process made of
\begin{enumerate*}[label=(\roman*)]
\item the sender's decision to initiate a relational event at a certain point in time, and
\item the sender's decision to address a specific receiver.
\end{enumerate*} 
The DyNAM decomposes event rates into two, sender-centred, components, i.e.,
\begin{equation}
\label{eq:stadtfeld2017interactions_2}
\lambda_{sr} (t) = \underbrace{\lambda_{s} (t)}_{\text{select sender $s$}} \times 
\underbrace{p \left(r \mid s \right)}_{\text{\parbox{2.5cm}{\centering sender $s$ \\[-5pt] addresses~receiver~$r$} }} \hspace{-8mm}.
\end{equation}
Similar to \citet{snijders2005models}, senders' event rates rates $\lambda_{s}$ are modeled through an exponential link function evaluating nodal attributes at individual and network levels,
\begin{equation}
\label{eq:stadtfeld2017interactions_3}
\lambda_{s} (t) = \lambda_0 \exp \, \left( \theta^\top x_{s} \right) \mathbbm{1}_{\left\{s\in \mathcal{R}(t) \right\}}.
\end{equation}
Following \citet{mcfadden1973conditional}, receiver choice is modeled via a multinomial distribution, i.e.,
\begin{equation}
\label{eq:stadtfeld2017interactions_5}
p \left(r \mid s, \mathcal{H}_{t} \right) = 
\myfrac{\exp \left\{ \beta^{\top} x_{sr} (t) \right\}}{ \sum_{r^\prime \in \mathcal{R}_{s} (t)} \exp \left\{ \beta^{\top} x_{sr^\prime}(t) \right\}}.
\end{equation} 
The timing distribution does not explicitly depend on the receiver characteristics and this typically sacrifices model fit over an actor-oriented interpretation of the model parameters.

The DyNAM has recently been extended with the incorporation of the DyNAM-i \citep{hoffman2020model}, which explains sequences of joining and leaving events in the context of group-based interactions. This extension captures 
\begin{enumerate*}[label=(\roman*)]
\item the specific nature of group conversations and interactions that typically occur in cliques; and
\item the need to align network modeling strategies with the increasing use of sensor technologies, such as Bluetooth or RFID badges, to detect collective interaction.
\end{enumerate*}

\citet{vu2017relational} exploited the separability of intensity functions by decomposing the stream of relational events into event times and event destinations. The separable sender intensity and receiver choice model adds more flexibility to the relational event framework allowing for the separation between senders and receiver effects, and not only between event weights and dyads as in \citet{brandes2009networks}.

\subsection{Network Covariates}
\label{subsec:network-covariates}

Given the general expression of the REM hazard in Eq.~\ref{eq:general-rem}, the drivers of the relational process $x_{st}(t)$ describe known, $\mathcal{H}_{t-}$ measurable, quantities quantifying endogenous or exogenous statistics of the sender, the receiver or both. 

\subsubsection{Endogenous VS Exogenous Covariates}
\label{subsubsec:endogenous-exogenous_network-covariates}

In statistical models for networks \citep[e.g.,][]{snijders2010introduction}, covariates are endogenous to the extent that they depend on past interaction. Covariates are exogenous when they depend on characteristic of single nodes (monadic covariates) or pairs or nodes (dyadic covariates). One example of endogenous covariate is reciprocity, while gender and geographical distance are exogenous covariates, representing monadic and dyadic characteristics, respectively. A basic selection of endogenous and exogenous covariates is displayed in Table~\ref{tab:netstats_formulae}. Other ad-hoc specifications, such as those based on exchange sequences in conversational analysis, are explained in \citet{butts2008relational}.

%
%
%
%
%
%
%
%
%
%

\begin{table}
		\caption{Basic menu of network covariates defined in the REM}
		\label{tab:netstats_formulae}
		\begin{center}
		\begin{adjustbox}{max width=0.9\textwidth}
		\begin{threeparttable}
\small
\begin{tabularx}{\textwidth}{
>{\raggedright \arraybackslash \hsize=0.25\hsize}C%
>{\centering \arraybackslash \hsize=0.30\hsize}C%
>{\centering \arraybackslash \hsize=0.45\hsize}C}%
\toprule
		Network Covariate &  Mechanism & Formula \\
		\bottomrule
		\noalign{\bigskip}
		Out-degree 
		& \begin{tikzpicture}[scale=0.45, baseline=(current bounding box.south)]
			\tikzset{every node/.style={draw=black, fill=tikz_white, shape=circle, inner sep=1.5pt, minimum size=0.28cm}}
			\node (a) at (-1.5,-1.5) [label={[label distance=-0.0cm]90:$a$}]{};
			\node (c) at (0.5,-1.5) [label={[label distance=-0.0cm]90:$s$}]{};
			\node (d) at (2,-0.5){};
			\node (e) at (2,-1.5) {};
			\node (f) at (2,-2.5) {};
			\draw[edge,<-,dashed] (a)--(c);
			\draw[edge,->] (c)--(d);
			\draw[edge,->] (c)--(e);
			\draw[edge,->] (c)--(f);
		\end{tikzpicture}
		& $\displaystyle \sum\limits_{a \neq s} N_{sa}(t^{-})$ \\ 
		\noalign{\medskip}
		\noalign{\medskip}
		Out-intensity
		& \begin{tikzpicture}[scale=0.45, baseline=(current bounding box.south)]
			\tikzset{every node/.style={draw=black, fill=tikz_white, shape=circle, inner sep=1.5pt, minimum size=0.28cm}}
			\node (a) at (-1.5,-1.5) [label={[label distance=-0.0cm]90:$a$}]{};
			\node (c) at (0.5,-1.5) [label={[label distance=-0.0cm]90:$s$}]{};
			\node (d) at (2,-0.5){};
			\node (e) at (2,-1.5) {};
			\node (f) at (2,-2.5) {};
			\draw[edge,<-,dashed] (a)--(c);
			\draw[edge,->, ultra thick] (c)--(d);
			\draw[edge,->, thin] (c)--(e);
			\draw[edge,->] (c)--(f);
		\end{tikzpicture}%
		& $ \myfrac{1}{\text{out-degree}_{s}(t)} \displaystyle \sum\limits_{a \neq s} \displaystyle \sum\limits_{i:~s_i=s,r_i=a} f \Big(t, t_{i}, \alpha \Big) $ \\
		\noalign{\medskip}
		\noalign{\medskip}
		In-degree
		& \begin{tikzpicture}[scale=0.45, baseline=(current bounding box.south)]
			\tikzset{every node/.style={draw=black, fill=tikz_white, shape=circle, inner sep=1.5pt, minimum size=0.28cm}}
			\node (a) at (-1.5,-1.5) [label={[label distance=-0.0cm]90:$a$}]{};
			\node (c) at (0.5,-1.5) [label={[label distance=-0.0cm]90:$s$}]{};
			\node (d) at (2,-0.5){};
			\node (e) at (2,-1.5) {};
			\node (f) at (2,-2.5) {};
			\draw[edge,->,dashed] (a)--(c);
			\draw[edge,<-] (c)--(d);
			\draw[edge,<-] (c)--(e);
			\draw[edge,<-] (c)--(f);
		\end{tikzpicture}%
		& $ \displaystyle \sum\limits_{a \neq s} N_{as}(t^{-}) $ \\
		\noalign{\medskip}
		\noalign{\medskip}
		In-intensity
		& \begin{tikzpicture}[scale=0.45, baseline=(current bounding box.center)]
			\tikzset{every node/.style={draw=black, fill=tikz_white, shape=circle, inner sep=1.5pt, minimum size=0.28cm}}
			\node (a) at (-1.5,-1.5) [label={[label distance=-0.0cm]90:$a$}]{};
			\node (c) at (0.5,-1.5) [label={[label distance=-0.0cm]90:$s,r$}]{};
			\node (d) at (2,-0.5){};
			\node (e) at (2,-1.5) {};
			\node (f) at (2,-2.5) {};
			\draw[edge,->,dashed] (a)--(c);
			\draw[edge,<-, ultra thick] (c)--(d);
			\draw[edge,<-, thin] (c)--(e);
			\draw[edge,<-] (c)--(f);
		\end{tikzpicture}%
		& $ \myfrac{1}{\text{in-degree}_{s}(t)} \displaystyle \sum\limits_{a \neq s} \displaystyle \sum\limits_{i:~s_i=a,r_i=s} f \Big(t, t_{i}, \alpha \Big)$ \\
		\noalign{\medskip}
		\noalign{\medskip}
		Repetition
		& \begin{tikzpicture}[scale=0.45, baseline=(current bounding box.south)]
			\tikzset{every node/.style={draw=black, fill=tikz_white, shape=circle, inner sep=1.5pt, minimum size=0.28cm}}
			\node (a) at (0,0) [label={[label distance=-0.0cm]90:$s$}]{};
			\node (b) at (2.5,0)[label={[label distance=-0.0cm]90:$r$}]{};
			\draw[edge,<-,dashed] (b.-210)--(a.30);
			\draw[edge,<-] (b.210)--(a.-30);
		\end{tikzpicture}%
		& $N_{sr}(t^{-})$\\
		\noalign{\medskip}
		\noalign{\medskip}
		Reciprocation
		& \begin{tikzpicture}[scale=0.45, baseline=(current bounding box.south)]
			\tikzset{every node/.style={draw=black, fill=tikz_white, shape=circle, inner sep=1.5pt, minimum size=0.28cm}}
			\node (a) at (0,0) [label={[label distance=-0.0cm]90:$s$}]{};
			\node (b) at (2.5,0)[label={[label distance=-0.0cm]90:$r$}]{};
			\draw[edge,<-, dashed] (b.-210)--(a.30);
			\draw[edge,->] (b.210)--(a.-30);
		\end{tikzpicture}%
		& $ N_{rs}(t^{-}) $ \\
		\noalign{\medskip}
		\noalign{\medskip}
		Transitive closure
		& \begin{tikzpicture}[scale=0.45, baseline=(current bounding box.south)]
			\tikzset{every node/.style={draw=black, fill=tikz_white, shape=circle, inner sep=1.5pt, minimum size=0.28cm}}
			\node (a) at (0,0) [label={[label distance=-0.0cm]180:$s$}]{};
			\node (b) at (2,0)[label={[label distance=-0.0cm]0:$r$}]{};
			\node (c) at (1,1.45) [label={[label distance=-0.0cm]90:$a$}]{};
			\draw[edge,->,dashed] (a)--(b);
			\draw[edge,->] (a)--(c);
			\draw[edge,->] (c)--(b);
		\end{tikzpicture}%
		& $ \displaystyle \sum\limits_{a \neq s, r} N_{sa}(t^{-}) N_{ra}(t^{-}) $ \\
		\noalign{\medskip}
		\noalign{\medskip}
		Cyclic closure
		& \begin{tikzpicture}[scale=0.45, baseline=(current bounding box.south)]
			\tikzset{every node/.style={draw=black, fill=tikz_white, shape=circle, inner sep=1.5pt, minimum size=0.28cm}}
			\node (a) at (0,0) [label={[label distance=-0.0cm]180:$s$}]{};
			\node (b) at (2,0)[label={[label distance=-0.0cm]0:$r$}]{};
			\node (c) at (1,1.45) [label={[label distance=-0.0cm]90:$a$}]{};
			\draw[edge,->,dashed] (a)--(b);
			\draw[edge,->] (c)--(a);
			\draw[edge,->] (b)--(c);
		\end{tikzpicture}%
		& $ \displaystyle \sum\limits_{a \neq s, r} N_{as}(t^{-}) N_{ra}(t^{-}) $ \\
		\noalign{\medskip}
		\noalign{\medskip}
		Sending balance
		& \begin{tikzpicture}[scale=0.45, baseline=(current bounding box.south)]
			\tikzset{every node/.style={draw=black, fill=tikz_white, shape=circle, inner sep=1.5pt, minimum size=0.28cm}}
			\node (a) at (0,0) [label={[label distance=-0.0cm]180:$s$}]{};
			\node (b) at (2,0)[label={[label distance=-0.0cm]0:$r$}]{};
			\node (c) at (1,1.45) [label={[label distance=-0.0cm]90:$a$}]{};
			\draw[edge,->,dashed] (a)--(b);
			\draw[edge,->] (a)--(c);
			\draw[edge,->] (b)--(c);
		\end{tikzpicture}%
		& $ \displaystyle \sum\limits_{a \neq s, r} N_{sa}(t^{-}) N_{ra}(t^{-}) $ \\
		\noalign{\medskip}
		\noalign{\medskip}
		Receiving balance
		& \begin{tikzpicture}[scale=0.45, baseline=(current bounding box.south)]
			\tikzset{every node/.style={draw=black, fill=tikz_white, shape=circle, inner sep=1.5pt, minimum size=0.28cm}}
			\node (a) at (0,0) [label={[label distance=-0.0cm]180:$s$}]{};
			\node (b) at (2,0)[label={[label distance=-0.0cm]0:$r$}]{};
			\node (c) at (1,1.45) [label={[label distance=-0.0cm]90:$a$}]{};
			\draw[edge,->,dashed] (a)--(b);
			\draw[edge,->] (c)--(a);
			\draw[edge,->] (c)--(b);
		\end{tikzpicture}%
		& $ \displaystyle \sum\limits_{a \neq s, r} N_{as}(t^{-}) N_{ar}(t^{-}) $ \\
		\noalign{\medskip}
		\noalign{\medskip}
		Node attribute
		& \begin{tikzpicture}[scale=0.45, baseline=(current bounding box.south)]
			\tikzset{every node/.style={draw=black, fill=tikz_color, shape=circle, inner sep=1.5pt, minimum size=0.28cm}}
			\node (c) at (0, 0) [label={[label distance=-0.0cm]90:$s,r$}]{};
		\end{tikzpicture}%
		& $x_{sr}(t)$ \\
		\noalign{\medskip}
		\noalign{\medskip}
		Node matching
		& \begin{tikzpicture}[scale=0.45, baseline=(current bounding box.south)]
			\tikzset{every node/.style={draw=black, fill=tikz_color, shape=circle, inner sep=1.5pt, minimum size=0.28cm}}
			\node (c) at (0, 0) [label={[label distance=-0.0cm]90:$s$}]{};
			\node (d) at (2.5, 0) [label={[label distance=-0.0cm]90:$r$}]{};
			\draw[edge,->,dashed] (c)--(d);
		\end{tikzpicture}%
		& $\text{dist} \left\{ x_{s}(t) = x_{r}(t) \right\}$ \\
		\noalign{\medskip}
		\bottomrule
	\end{tabularx}
\begin{tablenotes}[para, flushleft]
\small
\emph{Notes}. (Out/in)-(degree/intensity) statistics can refer to both senders and receivers. Node $a$ is a third (alter) trading counterpart of the sender-receiver pair $(s,r)$. The term $N_{sr}(t^{-})$ is the number of relational events flowing from sender $s$ to receiver $r$ right before time $t$, while $f(t, t_{i}, \alpha) = (t-t_{i})^{-\alpha}$ is the decay function accounting for the temporal relevance of previous relational events. In depicting network statistics, solid line arrows ($\rightarrow$) refer to past relational events, while dashed arrows ($\dashrightarrow$) indicate current relational events. 
\end{tablenotes}
\end{threeparttable}
\end{adjustbox}
\end{center}
\end{table}
				

An additional consideration refers to the \emph{hierarchy principle}, whereby lower-order interaction terms should always be included in the presence of higher-order interaction terms \citep{pattison2002neighborhood, wang2013exponential}. In the REM, for example, failing to account for heterogeneity of the senders and receivers may result in incorrect detection of triadic effects \citep{juozaitiene2022nodal}.

\subsubsection{Heterogeneity}
\label{subsubsec:heterogeneity}

There are two fundamental types of heterogeneity in event networks. Endogenous heterogeneity, or emergence, refers to the inherent stochasticity of the process itself combined with the dependence of future interactions on current ones. An example is \emph{virality}, whereby, for instance, a paper gets cited because it was cited many times before. 
In the REM, endogenous heterogeneity can be captured by endogenous covariates.

The second type of heterogeneity, extrinsic variation, is either observed, such as the prestige of the institutions of the authors of a paper, or latent, such as the quality of the work it represents. Latent extrinsic heterogeneity in the REM can be expressed by means of random effects. \citet{juozaitiene2022nodal} and \citet{uzaheta2023random} proposed mixed effect extensions of the REM, i.e.,
\[ 
\lambda_{sr}(t) = \mathbbm{1}_{\{(s,r)\in \mathcal{R}(t) \}} \: \lambda_0(t) \: \exp \left\{ \beta^\top x_{sr}(t)+ \gamma^\top z_{sr}(t) \right\},
\]
with dyadic covariates $z_{rs}(t)$ and $\gamma\sim N(0, \Sigma)$ the random effects. Estimation of the random effects variance can be done via Expectation Maximization \citep{dempster1972em} or Laplace approximations of the likelihood \citep{pinheiro2006mixed}, or in certain cases via a penalized zero order spline approach \citep{wood2017generalized}.


\cite{dubois2013stochastic} suggested a model that accounted for sender and receiver heterogeneity by means of a stochastic block structure. More recently, in an analysis of a communication network \citet{juozaitiene2022nodal} showed that incorporating random effects for both the sender and receiver enhances the model fit compared to model specifications that solely rely on endogenous degree-based statistics. Therefore, the inherent differences between individuals in the network drives part of the heterogeneity in the interactions. 

\subsubsection{Stratification}
\label{subsubsec:stratification}

Conceptually, stratification can be introduced either to model event streams in multiplex networks or to account for heterogeneity by specifying different baseline intensity functions for individual sets of dyads.  \cite{perry2013point, bianchi2022from}, for example, use sender-based stratification, effectively allowing each sender to have its own individual baseline hazard, i.e.,
\[
\lambda_{sr} (t) = \lambda_{0s}(t)\exp \left\{ \beta_{0}^{\top} x_{sr} (t) \right\}\mathbbm{1}_{\left\{(s, r)\in \mathcal{R}(t) \right\}}.
\]
Receiver-based stratification is defined in a similar fashion, and usually occurs when there is heterogeneity in those nodes that are repeatedly targeted as receivers. In citation networks, for example, groundbreaking articles or patents have very distinct, individual citation profiles, which makes a receiver-based baseline hazards an attractive option, given that they not have to be estimated individually \citep{filippi2023dream}.

\citet{juozaitiene2022non} proposed a stratified version of the REM, in which distinct baseline hazards are associated with distinct families of temporal network effects, such as reciprocity in its direct and generalized forms. Subsequent baseline hazard estimation reveals the tendency of some endogenous covariates to have very specific temporal effect-profiles.
\section{Estimation and Computation}
\label{sec:estimation-computation}

The fundamental information about a sequence of relational events $\{(t_i, s_i, r_i) : i=1, \ldots, n \}$ is contained in its likelihood function. For a REM, this function can be expressed as the product of the conditional generalized exponential event time densities and their associated multinomial relational event probabilities, i.e., 
\begin{equation}
\label{eq:full-lkl}
L(\beta) = \prod_{i=1}^{n} \sum_{(s,r)\in \mathcal{R}(t_i)} \lambda_{sr}(t_i) \exp\left\{-\sum_{(s,r)\in \mathcal{R}(t_i)}\int_{t_{i-1}}^{t_i}\lambda_{sr}(\tau)d\tau\right\} \myfrac{\lambda_{s_ir_i}(t_i)} {\sum_{(s,r)\in \mathcal{R}(t_i)}\lambda_{sr}(t_i)},
\end{equation}
where $\lambda$ is a function of $\beta$ and $\mathcal{R}(t)$ is the risk set, i.e., the set of all possible events that could have occurred at time $t$. 

Estimating the parameters of REMs by maximizing the full likelihood poses several challenges. The likelihood function is indeed a complex object that involves explicit integration across the unknown hazard function and sums over large risk sets. In this section, we will explore computational alternatives proposed to overcome the complexity of the full likelihood approach.

\subsection{Partial Likelihood Estimation}
\label{subsec:partial-likelihood-approaches}

\begin{figure}
\centering
\includegraphics[width=0.9\textwidth]{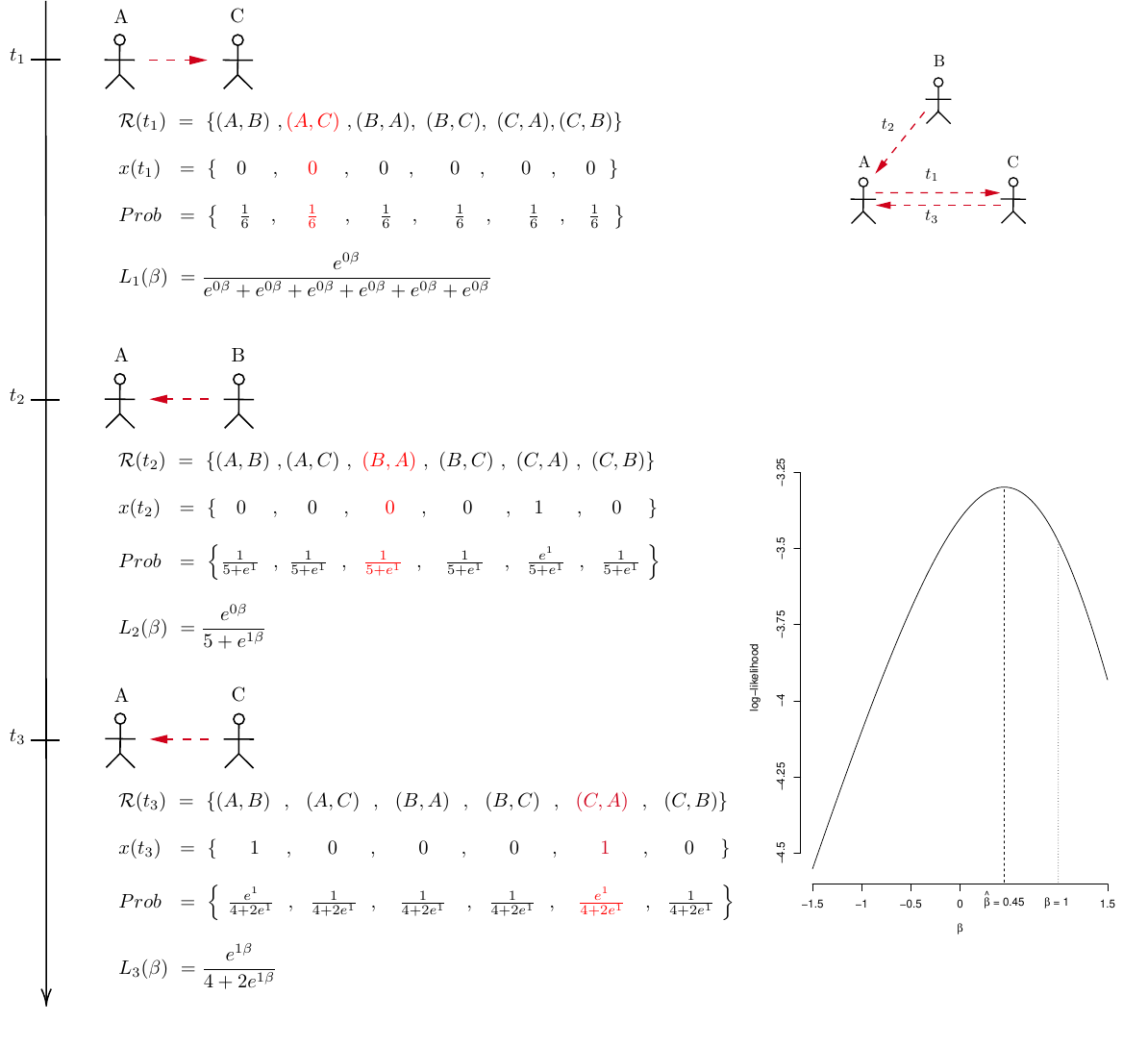}
\caption{
Example of a relational event process among three subjects $\{A,B,C\}$ based on the hazard function $\lambda_{sr}(t_i)=\lambda_{0}(t_i)\exp\left\{ \beta x(t_i) \right\}$, with $\beta=1$, whereby the event rate is influenced by the reciprocation statistic $x(t_i)$. The left-hand side of the figure outlines the unfolding of the sequence of relational events according to the true probabilities \emph{Prob}, mirrored in the graphical representation in the upper right. At each time point, the risk set $\mathcal{R}(t)$ consists of all six possible interactions between pairs of individuals. At time $t_1$, an interaction from $A$ to $C$ is observed, while all reciprocation covariates are zero, as no interaction has occurred yet. The probability of each event is determined using Eq.~\ref{eq:PL}, with partial likelihood $L_1(\beta)$. At time $t_2$, an interaction from $B$ to $A$ occurs. This event has partial likelihood $L_2(\beta)$, with one reciprocal non-event in the denominator. At $t_3$, the reciprocal interaction from $C$ to $A$ is observed, with partial likelihood $L_3(\beta)$. The overall partial likelihood $L_P(\beta)=\prod_{i=1}^{3}L_i(\beta)$ is maximized at $\hat{\beta}=0.45$, as shown in the lower-right side of the figure, close to the actual effect size $\beta=1$.
}	
\label{fig:rem}
\end{figure}

The proportional hazard model \citep{cox1984analysis} offers an attractive alternative to fully parametric models due to its absence of distributional assumptions regarding activity rates, which are then treated as nuisance parameters. It offers an effective simplification of the full REM likelihood through the application of the partial likelihood $L_P$ \citep{cox1975partial} to counting processes \citep{andersen1982cox} on network edges, which only involves multinomial event probabilities, i.e.,
\begin{equation}
\label{eq:PL}
L_P(\bm\beta) = \prod_{i=1}^{n} \left(\myfrac{\exp\left\{\beta^\top x_{s_ir_i}(t_i)\right\}}{\sum_{(s,r) \in \mathcal{R}(t_i)} \exp{\left\{\beta^\top x_{sr}(t_i)\right\}}}\right).
\end{equation}
This expression eliminates the unknown baseline hazard, resulting in a more adaptive representation of the underlying network dynamics, while being able to estimate the parameters in a straightforward way by maximizing $L_P(\beta)$. An example is given in Figure~\ref{fig:rem}. 

As \citet{butts2008relational} noted, the partial likelihood corresponds to the full likelihood when only the event orderings are known, but not the exact timings. However, the partial likelihood approach faces a limitation in large networks, as the risk set in its denominator tends to expand quadratically with the number of nodes.

\subsubsection{Risk Set Sampling}
\label{subsec:risk-set-sampling}

The primary constraints in modeling the partial likelihood from Eq.~\ref{eq:PL} are found in the risk set size. The idea of adopting subsampling strategies to approximate the partial likelihood was noted by \citet{butts2008relational}. \citet{vu2015relational} initially introduced a, more efficient, nested-case control sampling strategy \citep{borgan2015nested} to mitigate the computational complexity involved in estimating the partial likelihood. 

Nested case-control sampling consists of sampling from the current risk set $\mathcal{R}(t)$ according to some probability $\pi\{\cdot \mid \mathcal{R}(t)\}$ a set of non-events, or controls, for each event, or case. The sampled non-events together with the events are called the sampled risk set $\widetilde{\mathcal{R}}(t)$. \citet{borgan1995methods} show that the sampled partial likelihood $\widetilde{L}_P$, accounting for the sampling probabilities is a valid likelihood. When this probability is assumed to be random, i.e., $\pi\{\cdot \mid \mathcal{R}(t)\}= 1 / |\mathcal{R}(t)-1|$, $\widetilde{L}_P(\beta)$ reduces to the simplified form, i.e.,
\begin{align*}\label{eq:PL_sampled_risk}
\widetilde{L}_P(\bm\beta) &= \prod_{i=1}^{n} \left(\myfrac{\pi((s_i,r_i)|\mathcal{R}(t_i)) \exp \{{\beta^\top x_{s_ir_i}(t_i)\}}}{ \sum_{(s,r) \in \widetilde{\mathcal{R}}(t)} \pi((s,r)\mid \mathcal{R}(t_i)) \exp {\{\beta^\top x_{sr}(t_i)\}}}\right)\\
&= \prod_{i=1}^{n} \left(\myfrac{\exp \{{\beta^\top x_{s_ir_i}(t_i)\}}}{\sum_{(s,r) \in \widetilde{\mathcal{R}}(t)} \exp {\{\beta^\top x_{sr}(t_i)\}}}\right) \numberthis,
\end{align*}
where $\widetilde{\mathcal{R}}(t)$ is the sampled risk set. 

\citet{lerner2020reliability} employed nested case-control sampling to empirically showcase the efficiency of estimates on large networks, even when a limited number of non-events is sampled.

\subsubsection{Computational Aspects of Stratified Relational Event Models}
\label{subsec:computational_stratification}

\citet{perry2013point} proposed the first REM that introduced sender stratification to expedite calculations and combined it with a customized method for maximizing the log partial likelihood. \citet{vu2015relational} proposed a flexible stratification methods allowing for data structures with many types of nodes and events, then showing the scalability of their approach to large data sets. Combining the two approaches, \citet{bianchi2022from} proposed a sender-stratified REM for high-frequency data, using nested case-control sampling to update the risk set at each new event. This model specification has been tested in empirical applications using millions of financial transactions. \citet{filippi2023dream} proposed a receiver-stratified REM for the analysis of millions of patent citations, in which the hazard is modeled via smooth functions of the covariates using a spline approach.

\subsection{Baseline Hazard Estimation}
\label{subsec:baseline-hazard}

With the advancement of REMs, the prevailing method for their estimation is based on the partial likelihood method \citep{cox1975partial}, which treats the baseline hazard as a nuisance parameter. However, gaining insights into the temporal variations of the underlying event rates can be valuable for visualizing the baseline hazard.

Two common approaches to estimate the baseline hazard are the Breslow estimator \citep{breslow1972discussion} and the Nelson--Aalen estimator \citep{nelson1972hazard, aalen1978nonparametric}. Both estimate the cumulative hazard function only at the observed event times, and so does not capture the continuous nature of the underlying baseline hazard function. \citet{meijerink2022discovering} estimated the baseline hazard by assuming fixed baseline hazard rates within expanding windows. \citet{juozaitiene2022nodal} and \citet{juozaitiene2023analysing} improved smooth baseline hazard recovery by a spline-based approximation.

\subsection{Model Comparison and Diagnostic Tools}
\label{subsec:inference_model-comparison}

Traditional approaches, such as likelihood ratio tests, Akaike Information Criterion (AIC), and Bayesian Information Criterion (BIC), are widely adopted for comparing REMs \citep{foucault2014dynamic, pilny2016illustration}. Whereas model comparison methods are able to identify the best fitting model in a set of competing models, they do not indicate whether the fit is adequate. 

\citet{butts2017relational} propose an approach to model adequacy assessment based on deviance residuals and, so-called, surprise metrics. Similarly, \citet{meijerink2022discovering} suggest recall-based adequacy checking based on whether the observed events are in the top 5\% of dyads with the highest predicted rates.  
\citet{brandenberger2019predicting} proposes a procedure for model-based simulations of relational events. This method involves making predictions based on survival probabilities, which can then be used to generate new event sequences. In turn, by comparing the simulated event sequences with the original data, it becomes possible to assess whether the model can accurately replicate network characteristics.

Measures borrowed from survival analysis can also be used to assess the goodness of fit of REMs. \citet{juozaitiene2023analysing} propose using scaled Schoenfeld residuals to assess the proportional hazards assumption, deviance residuals to check for outliers and potential influential observations, and trends in martingale residuals to check for non-linear effects of the covariates. Different model specifications are compared according to their deviance explained through pseudo $R^2$ measures \citep{cox1989analysis}. \citet{boschi2023smooth} extended the martingale score process for evaluate the goodness of fit for fixed, smooth, and random effects. 

Guidelines on the statistical accuracy and precision of the REM are summarized in \citet{schecter2021power}, and defined by conducting experiments on simulated sequences of relational events, to which different sampling and scaling procedures have been applied. \citet{meijerink2022dynamic} showed that the accuracy and precision of REM estimates depend on the width of the selected temporal window. 

\subsection{Bayesian Estimation of the Relational Event Model}
\label{subsec:bayesian-estimation}

REMs can be fitted also using Bayesian approaches \citep{butts2008relational}. \citet{dubois2013hierarchical} extended the standard REM to incorporate multiple sequences, proposing a hierarchical model. \citet{mulder2019modeling} modeled multiple event sequences, estimating parameters that capture both within-sequence and between-sequence variations. This is particularly useful in multiplex networks, where multiple relational event sequences may be observed within the same network. \citet{vieira2022bayesian} introduced a Bayesian hierarchical model that enables inference at the actor level, providing valuable insights into the drivers influencing actors' preferences in social interactions. \citet{arena2021bayesian} and \citet{arena2023how} proposed different solutions for studying memory decay in REMs, showing that the Bayesian approach allows the estimation of short- and long-term memory effects on the model parameters in relational event sequences. 

\subsection{Tools for Analyzing Relational Event Models}
\label{subsec:tools_analysis}

There are limited specialized softwares fitting REMs. The \texttt{R}-based packages \texttt{relevent} and \texttt{informR} \citep{marcum2015constructing} are widely adopted, including all the REM features discussed in \citet{butts2008relational}. Another \texttt{R}-based option is the \texttt{rem} package \citep{brandenberger2018rem}, which allows the computation of endogenous covariates in signed one-, two-, and multi-mode event networks. Also the \texttt{remstats} package \citep{meijerink2022dynamic} assists empirical researchers in computing network covariates. It is typically used in combination with the \texttt{relevent} or \texttt{remestimate} packages for model estimation. A further \texttt{R}-based option is \texttt{goldfish}, which mainly supports DyNAM and is currently undergoing further development. 

\texttt{eventnet} \citep{lerner2020reliability} offers a reliable and scalable Java-based interface for the computation of endogenous and exogenous covariates that serve as inputs of proportional hazard regressions. \citet{bauer2021smooth} and \citet{fritz2021separable} showed that time-stamped relations data can be fitted through the well-established \texttt{R}-package \texttt{mgcv} \citep{wood2017generalized} for generalized additive models with penalized likelihoods after a proper data reorganization. 

One potential challenge for future studies is the development of a comprehensive package capable of computing network covariates under different sampling schemes while accommodating different model assumptions. Ideally, such a tool should be integrated within a suite of packages encompassing different estimation techniques and diagnostic tools as well.
\section{Applications of Relational Event Models}

\label{sec:emprical-applications}

Empirical studies adopting REMs span a wide range of disciplines. In this section, we classify these studies into established subjects, presented below in alphabetical order. Within each category, we offer an illustrative rather than exhaustive collection of research questions that have been investigated using REMs.

\subsection{Communication}
\label{subsec:communication}

Within the field of communication studies, broadly construed, REMs have been adopted to analyze conversational processes within and between organizations and teams as well as computer-mediated speeches. 

\citet{butts2008relational} and \citet{renshaw2022modeling} demonstrated the practical value of REMs in a study of organizational communication in the context of emergency management. REMs make it possible to specify covariates that capture basic conversational norms \citep{gibson2003participation, gibson2005taking}, such as the expectations of reciprocity in turn-taking.

\citet{leenders2016once} and \citet{quintane2016brokers} identified a number of challenges that hinder the identification of team dynamics and elaborated a REM-based analytic framework that supports a time-sensitive understanding of communication processes within teams. Similarly, \citet{quintane2013short}, \citet{pilny2016illustration}, and \citet{schecter2018step} revealed how REMs could be applied for studying the association between communication patterns and common indicators of process quality, coordination, and information sharing.

Computer-mediated communication is typically analyzed via two-mode REMs, which establish links between individuals and the situations they are involved in, such as the questions they answer in online Q\&A communities \citep{stadtfeld2011analyzing}, or problems they attempt to resolve in open-source online projects \citep{quintane2014modeling, tonellato2023microstructural}. In their study of communication instances in an online friendship network, \citet{foucault2014dynamic} adopt REMs to identify patterns of time dependence in data produced by online chats, focusing on how heterogeneous communication processes influence the creation, maintenance, and dissolution of communication ties over time.

\subsection{Ecology}
\label{subsec:ecology}

Studying behavior as sequences of relational events among animals promises to improve the understanding of key issues in behavioral ecology, such as how reciprocal giving and pro-social behavior emerges in small animal communities. Examples of this line of research include the study of \citet{tranmer2015using} on group interactions among captive jackdaws, with a focus on their food sharing habits, and among cows struggling with the introduction of unfamiliar members in their community \citep{patison_time_2015}.

REMs have been adopted for studying ecological niche invasions \citep{juozaitiene2023analysing} through the analysis of two-mode event networks linking invading species to territories. The relational event framework adopted in this study sheds light on potential risks associated with invasive species and develops insights into the ecological factors that may attract non-native species.

\subsection{Health and Healthcare}
\label{subsec:health-healthcare}

In health and healthcare research, REMs have been adopted in the study of social interaction in surgery rooms and inter-hospital patient transfers, with the aim of understanding how collaborations among healthcare units provides may or may not improve the quality of care.

\citet{pallotti2022lost} examined audio-visual recordings of task-related interactions among members of surgical teams to make sense of patterns of interpersonal communication among doctors and nurses organized around objects and technologies in the surgery room.

\citet{lomi2014quality} studied inter-hospital mobility in a small community of hospitals and found that decentralized patient sharing decisions ensure patients' access to higher-quality healthcare services. \citet{vu2017relational} showed that patient transfers are usually organized around small clusters of hospitals including reciprocated patient exchange. Studying collaborations among hospitals in a regional community in Southern Italy, \citet{amati2019some} explained that the generative mechanisms controlling change in event networks do not operate homogeneously and synchronously over time, but vary systematically over different days of the week. Similarly, \citet{zachrison2022influence} investigated the influence of characteristics such as reputation and institutional affiliation on the choice of destination hospital for emergency patients in the state of Massachusetts.

\subsection{Political Science}
\label{subsec:political-science}

Within the field of political science, REMs are frequently employed in their two-mode version. In the typical application, political actors are linked to social activities, such as participation in cosponsoring events or support expressed for claims. \citet{brandenberger2018trading} studied favor trading in congressional collaborations by examining the temporal dynamic of reciprocity, and found that the emergence of new collaboration clusters depends on the timing of mutual co-sponsorship. Due to the variety of actors involved in the political debate, \citet{haunss2022multimodal} adopted a multimodal extension of the DyNAM framework to study the political discourse around Germany's nuclear energy phase-out. This work identifies the potential discursive mechanisms that may have influenced the debate, and when they may have operated. 

REMs have occasionally been employed in criminology studies to investigate various aspects of offending behavior among individuals \citep{niezink2022things}. These studies have explored phenomena such as illegal drug exchange in online markets \citep{duxbury2021shining, duxbury2023network} and the replication tendency of gang violence \citep{gravel2023rivalries}. Overall, these analyses have demonstrated that criminal networks and personal attributes exert a substantial and temporally contingent influence on individual criminal acts.

\subsection{Sociology}
\label{subsec:sociology}

Because of a general familiarity with social network methods and models, sociology is perhaps the field where REMs have found more extensive application. 

Restricting the focus on reciprocity, \citet{kitts2017investigating} and \citet{lomi2021time} found that the mutual exchange of resources does not operate uniformly across different exchange regimes, time frames, and material settings defined, for instance, by the value of resources being exchanged \citep{zappa2021markets, bianchi2022from}.

\citet{lerner2017third} and \citet{lerner2020free} analyzed the emergence of status and hierarchies under conditions of extreme decentralization characterizing the Wikipedia crowdsourced project, in which independent volunteers interact by editing pages. In a study of individual editing behavior and collaboration/contention among Wikipedia editors, \citet{lerner2019team} examined the relation between team diversity, polarization, and productivity. 

In the sociology of education, REMs have been applied to study how the nature of learning environments affects students' outcomes. \citet{vu2015relational} studied educational experiences in massive open online courses by analyzing interactions between students with, and through an online learning interface.between students and their learning interface. \citet{dubois2013hierarchical} investigated the social dynamics of high school classrooms by considering how the individual propensity to share information is affected by factors such as seating arrangements, teaching style, or sequences of participation shifts in conversation among students and teachers.
\section{Open Issues and Challenges}

\label{sec:open-issues-challenges}

In this section we describe a number of open challenges in the context of relational event modeling. Although not exhaustive, it points to a number of interesting directions in which we expect significant progress in the near future. 

\subsection{Procedures for Assessing Goodness of Fit}
\label{subsec:gof-procedures}

A pressing issue in relational event modeling involves the absence of a comprehensive procedure for assessing goodness of fit. Methods have been proposed involving recall adequacy, and traditional residuals approaches.However, there exists no general consensus regarding a formal testing paradigm. In general, what seems to be missing is an approach to goodness of fit of REMs consistent with established auxiliary variable approaches developed for assessing the goodness of fit of ERGMs \citep{hunter2008goodness} and SAOMs \citep{lospinoso2019goodness}.

\subsection{Relational Big Data}
\label{subsec:massive_rem_data}

Inference for REMs encounters a computational bottleneck as the number of relational events and, specifically, the number of actors increases. This presents practical implications, as it limits the applicability of REMs to large-scale networks. Addressing this limitation is an essential unresolved matter for REMs.

Building upon the risk set sampling concepts introduced in \citet{vu2015relational} and \citet{lerner2020reliability}, \citet{filippi2023dream} have proposed a \emph{Stochastic Gradient Relational Event Additive Model} (STREAM) to analyze the network of patent citations in a dataset consisting of over 100 million events and 8 million nodes. By integrating case-control sampling with deep learning techniques, they have successfully achieved significant computational efficiency and real-time estimation of the model parameters.

\subsection{Current Developments of Relational Event Models}
\label{subsec:variation_of_rem}

\citet{fritz2022all} introduced a \emph{Relational Event Model for Spurious Events} (REMSE) to address the issue of events recorded through machine-coding errors, which can give rise to false positives and false negatives. The REMSE is presented as a robust tool suitable for studying relational events generated in potentially error-prone contexts. Its intensity is decomposed into two components: one associated with true events and the other with spurious events. This decomposition allows for a more accurate representation and understanding of the underlying dynamics in the presence of potentially erroneous data.

Within the realm of coauthorship networks, \citet{lerner2023micro} adapted the RHEM \citep{lerner2021dynamic} to settings where events have a measurable outcome, such as a performance measure. These outcomes can serve as additional explanatory variables in the RHEM or can be used as response variable. This extension, known as the \emph{Relational Hyperevent Outcome Model} (RHOM), implies that event rates and relational outcomes are determined by the same explanatory variables utilized in the RHEM.

Incorporating relational event dynamics into a latent space or latent clustering allows for novel hypotheses about drivers of network formation. \citet{dubois2013stochastic} combined ideas from stochastic block modeling and REMs by partitioning the node-set, where event dynamics within and between blocks evolve in distinct ways. \citet{matias_semiparametric_2018} developed a variational expectation-maximization method to estimate the latent groups. Moreover, combining REMs with latent space modeling allows the representation of actors as points in space, whose mutual distances drive the relational event process. \citet{artico_dynamic_2022} proposed a Kalman filter-based approach to estimate the trajectories of an actor in a latent space. An alternative approach is discussed in \citet{rastelli_continuous_2021}, where the likelihood of an event given the current latent positions is maximized by stochastic gradient descent.
\section{Concluding Remarks}
\label{sec:summary_conclusions}

Since their introduction fifteen years ago \citep{butts2008relational}, REMs have undergone considerable refinement \citep{brandes2009networks, perry2013point}, encouraged important extensions \citep{lerner2023relational}, and enabled development of substantive applications \citep{vu2015relational, vu2017relational, lerner2020free}. Progress on these various fronts contributed to establish relational event modeling as one of the most promising frameworks for the analysis of dynamic network data.

REMs add to the existing set of statistical models for the analysis of dynamic networks the possibility of using the information contained in the sequential order of social interaction events when social interaction events are transformed into network ties defined at more aggregate time scales. Conversation \citep{gibson2003participation}, financial transactions \citep{bianchi2022from}, technology-mediated communication \citep{butts2008relational}, problem-solving \citep{tonellato2023microstructural}, disaster management \citep{renshaw2022modeling}, medical emergencies \citep{zachrison2022influence}, are only few examples of processes where the sequential timing of relational events is essential for understanding the underlying observation-generating mechanisms. In situations characterized by comparable sequential constraints on social interaction, statistical models that assume the concurrency of network ``ties'' leave unresolved problems related to the fact that network mechanisms operate over different time frameworks, and are regulated by different time-clocks \citep{bianchi2022multiple}.

This review outlined the core properties and the mathematical underpinnings of REMs by tracking the development of the original model since its appearance in 2008. We devoted special attention to the challenges posed by the estimation of REMs, and discussed the computational approaches proposed to address the complexities of the original model and its successive variants. We emphasized the flexibility of the relational event modeling framework, which allows empirical specification to account for endogenous as well as exogenous covariates that may affect observed patterns of interaction. We discussed the various ways in which time may influence the impact that past events may have on future events --- an issue that may be considered an empirical feature of the data that should be accounted for, or an opportunity to develop theoretically inspired hypotheses about how time affects social interaction.

We intended our review to appeal to a broad audience comprising both empirically minded researchers confronting problems posed by the analysis of relational event data with complex temporal dependencies, and statisticians interested in the analytical opportunities offered by recent advances in dynamic stochastic models for social interaction. We expect that future progress in the modeling of relational events, and social networks more generally, will depend on the extent to which members of these communities will continue to discover areas of intersection for their interests.

\section*{Acknowledgments}
We are grateful to J\"urgen Lerner for his comments on earlier versions of this manuscript, Carter Butts for his apposite advice, and to an anonymous reviewer for their careful reading of our work and helpful editorial advice. The authors acknowledge funding from the Swiss National Science Foundation (SNSF) through grant number 192549.

\bibliography{00_references}  

\end{document}